\documentclass[12pt]{elsart}
\usepackage{epsf}
\usepackage{epsfig} 
\usepackage{amssymb}
\usepackage{amsmath}
\usepackage[dvips]{color}
\usepackage{multicol}
\newcommand{\bm}[1]{{\mbox{\boldmath$#1$}}}
\def\Quadrat#1#2{{\vcenter{\hrule height #2
\hbox{\vrule width #2 height #1 \kern#1
\vrule width #2}\hrule height #2}}}
\def\dAl{\mathop{\kern 1pt\hbox{$\Quadrat{8pt}{0.4pt}$} \kern1pt}}
\begin{document}
\begin{frontmatter}
\baselineskip=12pt
\title{The Post-Newtonian Treatment of the VLBI Experiment on September 8, 2002}
\author{Sergei M. Kopeikin}
\address{Department of Physics and Astronomy, University of
Missouri-Columbia, Columbia, MO 65211, USA}\maketitle
\begin{abstract}
Gravitational physics of VLBI experiment conducted on September 8, 2002 and dedicated to measure the speed of gravity (a fundamental constant in the Einstein equations) is treated in the first post-Newtonian approximation. Explicit speed-of-gravity parameterization is introduced to the Einstein equations to single out the retardation effect caused by the finite speed of gravity in the relativistic time delay of light, passing through the variable gravitational field of the solar system. The speed-of-gravity 1.5 post-Newtonian correction to the Shapiro time delay is derived and compared with our previous result obtained by making use of the post-Minkowskian approximation. We confirm that the 1.5 post-Newtonian correction to the Shapiro delay depends on the speed of gravity $c_g$ that is a directly measurable parameter in the VLBI experiment.
\end{abstract}
\maketitle
\end{frontmatter}
\section{Introduction}
The gravitational VLBI experiment for measuring the relativistic effect of propagation of gravity field of orbiting Jupiter was conducted on September 8, 2002 by the 
National Radio Astronomical Observatory (USA) and the Max Plank Institute for Radio Astronomy (Germany). The idea of the experiment was proposed by Kopeikin \cite{2} who had 
noted that an arbitrary-moving gravitating body deflects light not instantaneously but with retardation caused by the finite speed of gravity propagating from the body to the light ray. This is because positions of the gravitating bodies are connected to the light-ray particle by the gravity null cone which is defined by the equation for the retarded time in the gravitational {\it Li\'enard-Wiechert}  potentials being solutions of the Einstein equations.  The goal of the experiment was to confirm this prediction by making use of the close celestial alignment of Jupiter and the quasar J0842+1835. The goal of this paper is to make use of the parameterized post-Newtonian technique to reveal how the gravity propagation affects the light-ray trajectory. This approach allows us to confirm our previous theoretical results \cite{2} by looking at the problem from different position.

\begin{figure}
\centerline{\psfig{figure=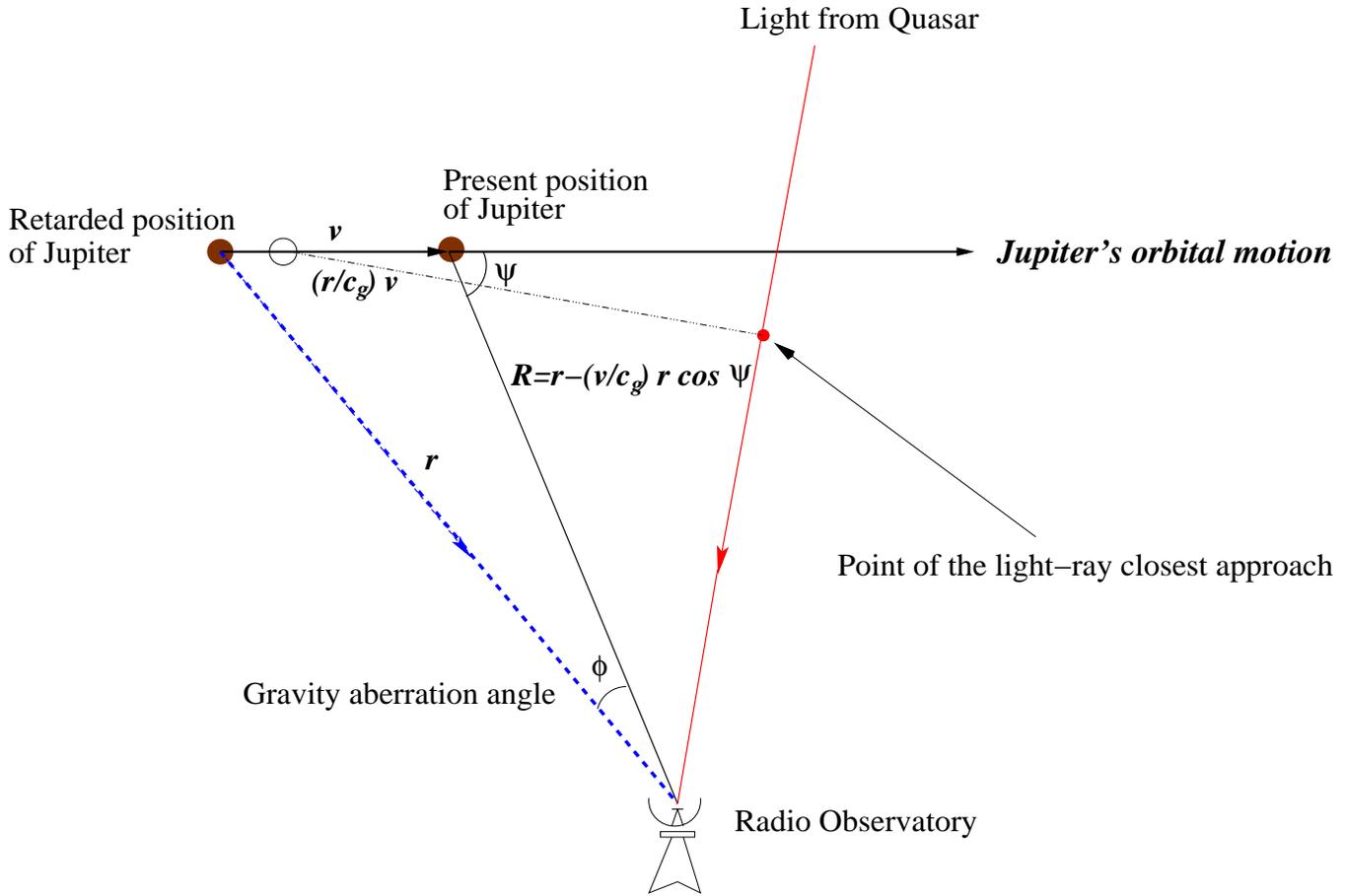,angle=0,height=12cm,width=18cm}}\vspace{1cm}
\caption{Light propagates with speed $c$ from the source of electromagnetic waves (quasar) to radio observatory. A massive body (Jupiter) orbits the Sun with speed ${\bm v}$. Perturbation of 
gravity field produced by the orbital motion of Jupiter propagates from Jupiter to the light ray with the speed $c_g=c$. In particular, the gravity-field perturbation
arrives to the point of observation of radio signal at time $t_1$. In
order to calculate the delay in the time of
arrival of light from the quasar to observer, position of Jupiter must be calculated at the retarded
time $s_1=t_1-r/c_g$ irrespectively of the direction of propagation of light \cite{2}. The present position of Jupiter at the time $t_1$ is displaced with respect to its retarded position
at the distance $(v/c_g)r \cos\psi$ and does not play any role in
the calculation of the time delay of light. Sun and Earth produce similar effect. \label{fig1}}
\end{figure}

The problem is formulated as follows. Light ray is emitted by a quasar (QSO J0842+1835) at the time $t_0$ and moves to the network of VLBI stations located on the Earth. As light moves, it passes through the variable gravitational field of the solar system, created by the orbital motion of Sun and other massive planets, and is received by the first and second VLBI stations at the times $t_1$ and $t_2$ respectively. Gravitational field of the solar system causes delay in the time of propagation of radio signals -- the effect discovered by \cite{6} in static gravitational field approximation. We noted \cite{2} that the present-day accuracy of phase-reference VLBI measurements is sufficient to detect a 1.5 post-Newtonian\footnote{The first 1.0 post-Newtonian approximation is of order $c^{-2}$, and 1.5 post-Newtonian correction to the Newtonian gravity is of order $c^{-3}$. } correction to the Shapiro time delay incorporated implicitly into position of the light-ray deflecting body (see Fig.\ref{fig1}) which must be taken at the retarded instant of time related to the time of observation by the equation of the gravity null-cone. We solved the Einstein-Maxwell system of differential equations and calculated this correction for Jupiter by making use of the post-Minkowskian approximation \cite{2} with the retarded {\it Li\'enard-Wiechert} solutions of the Einstein equations. The retardation is caused by the finite speed of propagation of gravity $c_g=c$ which, as we have revealed in \cite{2}, enters the 1.5 post-Newtonian correction to the Shapiro time delay. Hence, we concluded that the VLBI experiment on September 8, 2002 is sensitive to the effect of the {\it propagation of gravity} and can be used to measure its speed.

Our primary paper \cite{2} has given no particular details on the parameterization of the gravity propagation effect the amplitude of which was characterized by the parameter $\delta$. For this reason the physical meaning of the gravity propagation effect was not understood adequately by some researchers \cite{1,w-astro}. By scrutiny inspection we found that \cite{1} made an erroneous use of insufficient 1.0 post-Newtonian (static) approximation to discuss the high-order 1.5 post-Newtonian (dynamic) correction to the Shapiro time delay while \cite{w-astro} gave analysis of the problem of propagation of light ray in the time-dependent gravitational field of the solar system on the basis of the PPN formalism which was not sufficiently elaborated to tackle properly the 1.5 post-Newtonian approximation to general relativity (see appendix \ref{a2}). 
 
In this paper we extend our theoretical analysis of the VLBI experiment by proving that the parameter $\delta=c/c_g-1$, where $c$ is the speed of light and $c_g$ is the speed of gravity. Parameter $\delta$ is a measurable quantity and its determination will provided us with the numerical estimate of the magnitude of the propagation of gravity effect in terms of the speed of gravity $c_g$. The spectrum of plausible values of $c_g$ ranges from $c_g=c$ in general relativity to $c_g=\infty$ as advocated by \cite{tvf}. The VLBI experiment gives $c_g/c=1.06\pm 0.21$ 
\cite{apj}. The values $c_g<c$ however are unlikely and ruled out, for example, by observations of cosmic rays \cite{mn}. 

The paper is organized as follows. In section 2 we discuss the linearized Einstein equations and the speed-of-gravity parameterization that is used in section 3 for derivation of the post-Newtonian metric tensor. Section 4 deals with the post-Newtonian time delay formula for light propagation. The effect of retardation of gravity in the gravitational time delay is explained in section 5. The differential VLBI time delay is derived in section 6 where we prove that our previous post-Minkowskian formulation of the VLBI delay is identical with the post-Newtonian presentation. Section 7 contains discussion of the basic results of the present paper.

Roman indices run from 1 to 3 while Greek indices run from 0 to 3. Repeated indices assume the Einstein summation rule. Indices are raised and lowered with the Minkowski metric $\eta_{\alpha\beta}={\rm diag}(-1,+1,+1,+1)$. The Kroneker symbol $\delta^\alpha_\beta={\rm diag}(1,1,1,1)$. We denote coordinates $x^\alpha=(ct,x^i)$ and partial derivatives $\partial_\alpha\equiv(c^{-1}\partial/\partial t, \partial/\partial x^i)$. Partial derivative associated with $c_g$ is denoted as $\eth_\alpha\equiv(c_g^{-1}\partial/\partial\tau, \partial/\partial x^i)$, where $\tau$ relates to time $t$ via Eq. (\ref{1}).

We use the boldface italic letters to denote spatial vectors, e.g, ${\bm a}\equiv a^i=(a^1,a^2,a^3)$. The dot between two spatial vectors is the Euclidean scalar product, ${\bm a}{\bm\cdot} {\bm b}=a^1b^1+a^2b^2+a^3b^3$. The cross between two spatial vectors denotes the Euclidean vector product, ${\bm a}\times{\bm b}\equiv \varepsilon_{ijk}a^jb^k$, where $\varepsilon_{ijk}$ is the Levi-Civita symbol such that $\varepsilon_{123}=+1$. 

\section{Linearized Einstein Equations and Their Parameterization}

We make use of the following decomposition of the metric tensor $g_{\mu\nu}$ 
\begin{equation}
\label{met}
\sqrt{-g}g^{\mu\nu}=\eta^{\mu\nu}-\gamma^{\mu\nu}\;,
\end{equation}
where $g={\rm det}\bigl[g_{\mu\nu}\bigr]$ and $\gamma^{\mu\nu}$ is the gravitational field perturbation. When $c_g=c$ the linearized Einstein equations in harmonic gauge, $\partial_\alpha\gamma^{\alpha\beta}=0$, read
\begin{eqnarray}
\label{ee}
\left(-\frac{1}{c^2}\frac{\partial^2}{\partial t^2}+{\bm\nabla}^2\right)\gamma^{\alpha\beta}(t, {\bm x})&=&-{16\pi G\over
c^4}\, T^{\alpha\beta}(t, {\bm x})\;,
\end{eqnarray}
where ${\bm\nabla}^2\equiv\partial^i\partial_i$ is the Laplace operator, and $T^{\alpha\beta}(t, {\bm x})$ is the stress-energy tensor of matter, and $G$ is the universal gravitational constant. 

Special relativity postulates that in every inertial frame, there is a finite universal limiting speed $c$. Experimentally, the limiting speed $c$ is equal to the speed of light $c_l$ in vacuum with amazing precision \cite{cg}. General relativity is a theory of gravitational field propagating with some speed $c_g$ that was equated by Einstein to the fundamental speed $c$ which appear in the wave operator in Eq. (\ref{ee}) while the right side of these equations contains a constant $16\pi G/c^4$ describing the coupling of gravity with matter. In what follows we intend to build up the parameterized version of the linearized Einstein equations (\ref{ee}) with the speed of gravity $c_g$ in such a way that allows to get both the Newtonian and special relativity limits simultaneously, i.e., $c_g\rightarrow\infty$ while $c={\rm const.}$ We emphasize, however, that our approach is just the first step in this direction and further development is required. 

Because in general relativity the speed of gravity $c_g$ is denoted by the same letter $c$ it is difficult (but possible) to discern the effects associated with $c_g$ from those associated with the fundamental speed $c$. Thus, a theoretical framework is needed to unravel $c_g$ from $c$ in the experimental gravitational physics. We shall develop in this paper a parameterized approximation to general relativity which allows to segregate $c_g$ from $c$ in the experiments conducted in weak gravitational fields. Development more advanced (non-linear) formalism goes beyond the scope of this paper. Such formalism will be required, for example, for setting experimental limits on the speed of gravity from binary pulsar timing \cite{wt}. 

In what follows, we shall keep notation for the speed of gravity and light as $c_g$ and $c$ respectively, and consider the speed of gravity $c_g$ as a parameter which can range from $c$ to $\infty$. Speed of gravity characterizes temporal variation of gravitational field and the pace of its propagation. This implies that partial time derivatives of the metric tensor perturbation $\gamma^{\alpha\beta}$ in the left side of the Einstein equations (\ref{ee}) must be parametrically associated with the speed $c_g$. This will lead to the wave operator depending explicitly on the speed of gravity $c_g$ considered as a parameter.

However, if one holds the time coordinate $t$ fixed for different values of the parameter $c_g\not=c$ it would change the numerical value of the time coordinate $x^0$ of one and the same event and mix up the order of space-time points. For this reason, we will keep numerical value of the time coordinate $x^0=ct$ to be the same while changing $c\rightarrow c_g$.
This assumes that the smooth transition from $c$ to $c_g$ in the left part of Eqs. (\ref{ee}) is to be accompanied with the introduction of a new time variable, $\tau$, which must scale simultaneously with the parameter $c_g$ in such a way that equality $c_g\tau=ct$ is preserved.

In other words, the variable $\tau$ must relate to time $t$ as follows 
\begin{eqnarray}
\label{1}
\tau&=&\epsilon t\;,
\end{eqnarray} 
where $\epsilon\equiv c/c_g=1+\delta$ is a dimensionless parameter that provides the speed-of-gravity parameterization of Eqs. (\ref{ee}) and is used in their post-Newtonian expansion to keep track of the orders in the post-Newtonian expansion.
As one will see later, the time $\tau$ parameterizes world lines of massive bodies comprising a gravitating system. Rescaling of the time variable means that the system with $\epsilon=1$ is doing at time interval $\Delta T$ what the system for any other $\epsilon\not=1$ is doing at time $\Delta T/\epsilon$, i.e. the faster gravity propagates $(\epsilon\rightarrow 0)$ the shorter the time scale $\tau$. In other words, we are approaching to the Newtonian limit along the sequence of solutions of Eqs. (\ref{ee}) parameterized by the sliding parameter $\epsilon$ but fixed value of $x^0=ct$. The Newtonian limit is given by $\epsilon =0$, and in the limit $\epsilon=1$ we restore the linearized Einstein equations (\ref{ee}). 

The left side of Eqs. (\ref{ee}) is parameterized by the speed of gravity after making use of the replacements: $c\rightarrow c_g$, $t\rightarrow\tau$, and $\gamma^{\alpha\beta}(t, {\bm x})\rightarrow\gamma^{\alpha\beta}(\tau, {\bm x})$. This is, however, not enough because, if one wants to keep the form of the Eq. (\ref{ee}) be formally the same as the original Einstein equations in the harmonic coordinates, the gauge conditions must be re-written as $\eth_\alpha\gamma^{\alpha\beta}(\tau, {\bm x})=0$. Since the gauge conditions are directly related to the equations of motion of gravitating matter \cite{ll}, their replacement implies that the  stress-energy tensor must be replaced as well: $T^{\alpha\beta}(t,{\bm x})\rightarrow \Theta^{\alpha\beta}(\tau,{\bm x})$, where
\begin{equation}
\label{w}
\Theta^{00}= T^{00}(\tau,{\bm x})\;,\qquad \Theta^{0i}= \epsilon\, T^{0i}(\tau,{\bm x})\;,\qquad \Theta^{ij}= \epsilon^2\, T^{ij}(\tau,{\bm x})\;.
 \end{equation}
 This law of re-scaling of the stress-energy tensor is chosen by the requirement that the Newtonian limit is obtained when $\epsilon\rightarrow 0$ while the speed of light $c$ is constant. As for the formal structure of $T^{\alpha\beta}(\tau,{\bm x})$ we assume that it is the same as in special relativity. Equations of motion for $\Theta^{\alpha\beta}(\tau,{\bm x})$ follow from the gauge conditions and are given by $\eth_\alpha\Theta^{\alpha\beta}=0$.
 
Finally, the parametric form of the linearized equations (\ref{ee}) in the parameterized harmonic gauge, $\eth_\alpha\gamma^{\alpha\beta}(\tau, {\bm x})=0$,  is
\begin{eqnarray}
\label{gfe}
\left(-{1\over c_g^2}{\partial^2\over\partial
\tau^2}+{\bm\nabla}^2\right)\gamma^{\mu\nu}(\tau, {\bm x})&=&-{16\pi G\over
c^4}\, \Theta^{\mu\nu}(\tau, {\bm x})\;.
\end{eqnarray}
We emphasize once again that it would be illogical to obtain Eq. (\ref{gfe}) from the Einstein equations (\ref{ee}) with a simple replacement $c\rightarrow c_g$ without changing the time coordinate $t\rightarrow\tau$. An attempt to leave the time coordinate unchanged would violate the order of events for different values of the sliding parameter $c_g$  and lead to different treatment of the problem of light propagation (see appendix \ref{a2}).

Smooth transformation of the linearized equations (\ref{ee}) of general relativity to the sequence of parametric space-times allows us to keep two speeds, $c$ and $c_g$, to be formally different for each value of $\epsilon\not=1$. The space-time parameterization is introduced via 
Eqs. (\ref{1})--(\ref{gfe}) which can be understood as mapping of the physical metric  $g_{\alpha\beta}(t,{\bm x})$ $(\epsilon=1)$ into the sequence of solutions $g_{\alpha\beta}(\tau=\epsilon t,{\bm x})$ of the linearized equations (\ref{gfe}), where $\epsilon$ is a parameter along the sequence. This sequence of solutions is generated for each value of $\epsilon$ by the parametric stress-energy tensor of matter $\Theta^{\alpha\beta}(\tau,{\bm x})$ which is locally conserved. The mapping rule for connecting of identical events, having the same spatial coordinates on different slices of the parametric space-time manifold, is $c_g\tau=ct$.

\section{The Post-Newtonian Metric Tensor} 

Let us work in the reference frame of the solar system with the origin located at its barycenter. 
Let $T^{\alpha\beta}$ be the stress-energy tensor of massive
point-like particles \cite{ll}
\begin{eqnarray}
\label{3a} T^{\mu\nu}(\tau, {\bm x})&=&\sum_{a=1}^N
M_a \gimel_a^{-1}(\tau)\,u_a^\mu(\tau)\,
u_a^\nu(\tau)\,
\delta^{(3)}\bigl({\bm x}-{\bf
x}_a(\tau)\bigr)\;,
\end{eqnarray}
where the index $a=1,2,...,N$
enumerates gravitating bodies of the solar system, $M_a$
is the (constant) rest mass of the $a$th body,
${\bm x}_a(\tau)$ are time-dependent spatial coordinates of the $a$th body, 
${\bm v}_a(\tau)= d{\bm x}_a(\tau)/d\tau$ is the orbital
velocity of the $a$th body, $u_a^\alpha=\gimel_a (c,\,{\bm v}_a)$ is the
four-velocity of the $a$th body, $\gimel_a=\bigl(1-v_a^2/c^2\bigr)^{-1/2}$,
and $\delta^{(3)}({\bm x})$ is a 3-dimentional Dirac's delta-function.

Solution of Eq. (\ref{gfe}) can be found
in terms of the retarded {\it Li\'enard-Wiechert} tensor potentials \cite{5}, \cite{km}
\begin{eqnarray}
\label{4a} 
\gamma^{\mu\nu}(\tau,{\bm x})\sim {4G\over
c^4}\sum_{a=1}^N\;\frac{M_a\, u_{a}^\alpha(s)\, u_{a}^\beta(s)}
{r_a(s)-c_g^{-1}\,{\bm v}_a(s)\cdot {\bm r}_a(s)}\;,
\end{eqnarray}
which depend on the retarded time $s=s(\tau,{\bm x})$ determined for each body as a
solution of the gravity null-cone equation \cite{2}
\begin{equation}
\label{rt} s=\tau-{r_a(s)\over c_g}\;,
\end{equation}
connecting position of $a$th body to the field point $(\tau,{\bm x})$, and ${\bm r}_a(s)={\bm x}-{\bm x}_a(s)$, $r_a(s)=|{\bm r}_a(s)|$. Notice that the retarded time $s$ is different for each body as it depends on the body's position ${\bm x}_a$.  

The post-Newtonian form of the metric tensor, used in the subsequent calculation of a light-ray propagation, is obtained after making use of the Taylor expansion of all functions in the left side of Eq. (\ref{4a}) about the present instant of time $\tau$. It yields  
\begin{eqnarray}
\label{mt1}
\gamma^{00}\left(\tau,{\bm x}\right)&=&4\,\sum_{a=1}^N {GM_a\over
c^2 R_a}+O\left(c^{-4}\right)\;,\\\label{mt2}
\gamma^{0i}\left(\tau,{\bm x}\right)&=&4\epsilon\sum_{a=1}^N {GM_a \over
c^3 R_a}\,v^i_a(\tau)+O\left(c^{-5}\right)\;,\\\label{mt3} 
\gamma^{ij}\left(\tau,{\bm x}\right)&=&4\epsilon^2\sum_{a=1}^N {GM_a\over
c^4 R_a}\,v^i_a(\tau) v^j_a(\tau)+O\left(c^{-5}\right)\;, 
\end{eqnarray}
where $R_a=|{\bm R}_a|$, ${\bm R}_a=R_a^i=x^i-x_a^i(\tau)$.
Let us note that in the Newtonian limit $c_g\rightarrow\infty$ ($\epsilon\rightarrow 0$) so that the post-Newtonian metric tensor (\ref{mt1})--(\ref{mt3}) is reduced to the simple case with $\gamma^{00}$ as the only significant gravitational field perturbation. This confirms the standard point of view that gravity propagates instanteneously in the Newtonian theory and is described by a single `scalar' function. It is worth noticing that $\gamma^{\mu\nu}$ directly relates to the metric perturbations $h_{\mu\nu}=g_{\mu\nu}-\eta_{\mu\nu}=\gamma_{\mu\nu}(\tau,{\bm x})-(1/2)\eta_{\mu\nu}\gamma^\alpha_{\;\;\alpha}(\tau,{\bm x})$. In the limit $\epsilon=0$ one has $h_{00}=1/2\gamma_{00}$ and $h_{ij}=1/2\delta_{ij}\gamma_{00}$. Thus, integration of the light ray propagation equation in this limit will yield exactly the Shapiro time delay for the gravitational field propagating instanteneously with infinite speed.  

Calculation of the post-Newtonian limit (\ref{mt1})--(\ref{mt3}) indicates that our parameterized model of the gravitational equations (\ref{gfe}) can be interpreted as having the following values of the PPN parameters: $\beta=\gamma=1$, $\alpha_1=8\delta$, $\alpha_2=\delta(\delta+2)$, and all other PPN parameters are zero. We emphasize however that this interpretation makes sense only if one considers the time variable $\tau$ as physical. However, in our formalism time $t$ is physical and coincides with $\tau$ for $\epsilon=1$. It should be also understood that the velocity ${\bm v}_a(\tau)=d{\bm x}_a(\tau)/d\tau$ of the gravitating bodies is numerically equal to the physical velocity ${\bm v}_a(t)=d{\bm x}_a(t)/dt$ in the approximation of uniform and rectilinear motion which we shall use later for analysis of the time delay \footnote{Indeed, if one considers ${\bm x}_a(\tau)=a+b\tau$ with constant $a$ and $b$, then ${\bm x}_a(t)=a+bt$ and for any $\epsilon\not=1$ one has ${\bm v}_a(\tau)={\bm v}_a(t)=b$. In other words, it is meaningless to define velocity as $d{\bm x}_a/dt$ or $d{\bm x}_a/d\tau$ without explicit indication of the argument of the function ${\bm x}_a$.}. This remark is important to avoid misinterpretation of the velocity ${\bm v}_a(\tau)$ in Eq. (\ref{aber}).

The parameterized post-Newtonian (PPN) formalism \cite{9} predicts
that certain preferred frame (Lorentz-invariance violating) effects, which are proportional to the
ratio $c/c_g$, would arise in case if $c_g\not= c$. These effects are
characterized by the product of either of two parameters $\alpha_1$, $\alpha_2$ with the speed ${\bm w}$ of the
solar system with respect to the preferred frame. Under assumption that  
the isotropy of the cosmic microwave background radiation (CMBR)
defines the preferred frame, a limit has been
established on $\alpha_1\le 10^{-4}$ \cite{9}.  But, we do not know
if the CMBR does, in fact, define the preferred coordinate system. For example,
future observations of relic gravitational wave background
(GWB) could lead to the frame moving somehow differently with
respect to the CMBR frame, since it is assumed that the GWB was formed in the very early universe, long before the
CMBR decoupled from matter. Furthermore, modern multiconnected cosmologies challenge the cosmological Copernican Principle and indicates various possibilities for the preferred frame be not coinciding with the CMBR frame \cite{blev}.
We conclude that the existing limit on
$\alpha_1$ is uncertain until we can
confidently determine the preferred reference frame and the speed ${\bm w}$ of the solar system with respect to it from laboratory experiments measuring violation of the local Lorentz-invariance. 

\section{The Post-Newtonian Time Delay}

In general relativity light propagates along null geodesics of the metric tensor $g_{\alpha\beta}(\tau, {\bm x})$. In our parameterized approach to general relativity we want to keep the speed of light equal to $c$ for any value of the parameter $\epsilon$. This assumes that the equations of light propagation must be ordinary null geodesics expressed in terms of the Christoffel symbols approximated as
\begin{equation}
\label{ch}
\Gamma^\alpha_{\mu\nu}=\frac{1}{2}\eta^{\alpha\beta}\left(
\partial_\nu h_{\beta\mu}+\partial_\mu h_{\beta\nu}-
\partial_\beta h_{\mu\nu}\right)\,,
\end{equation}
where $h_{\mu\nu}=\gamma_{\mu\nu}(\tau,{\bm x})-(1/2)\eta_{\mu\nu}\gamma^\alpha_{\;\;\alpha}(\tau,{\bm x})$, and the partial derivative $\partial_\alpha=(c^{-1}\partial/\partial t, \partial/\partial x^i)$ does not depend on $\epsilon$. The speed of gravity $c_g$ appears in equations of light geodesics and, consequently in the gravitational time delay, {\it explicitly} through the numerical factors in the components of the metric tensor perturbations, and {\it implicitly}, through the dependence of coordinates ${\bm x}_a(\tau)$ of gravitating bodies on time $\tau=\epsilon t=(c/c_g)t$. 

Because the speed of light is held constant we must use the coordinate time $t$ to parameterize light-ray trajectory while the time $\tau$ is used for parameterization of the world-lines of gravitating bodies deflecting the light ray. 
The undisturbed propagation of light in the absence of gravity is a straight line
\begin{equation}
\label{und}
x^i_N(t)=x^i_0+ck^i(t-t_0)\;,
\end{equation}
where $t_0$ is time of emission, $x^i_0$ is position of the source of light at time $t_0$, $k^i$ is the unit vector in the direction of propagation of light from the source of the light to observer. 

The post-Newtonian equation of light geodesic parameterized by coordinate time $t$ reads \begin{equation}
\label{lg}
{d^2x^i\over dt^2}=c^2k^\mu k^\nu\left(k^i\,\Gamma^0_{\mu\nu}-\Gamma^i_{\mu\nu}\right)\;,
\end{equation}
where $k^\mu=(1,k^i)$ is the four-vector of the unperturbed light-ray path, 
the Christoffel symbols are given by Eq. (\ref{ch}) and should be taken on the unperturbed light-ray trajectory (\ref{und}).

Double integration of Eq. (\ref{lg}) yields time of propagation of light from the point $x_0^i$ to $x^i_1\equiv x^i(t_1)$
\begin{equation}
\label{qer}
t_1-t_0={1\over c}|{\bm x}_1-{\bm x}_0|+\Delta(t_1,t_0)\;,
\end{equation}
where
\begin{equation}
\label{aa}
\Delta(t_1,t_0)={c\over 2}\;k^\mu k^\nu k^\alpha\int_{t_0}^{t_1}dt\int_{-\infty}^t d\zeta \Bigl[{\partial_\alpha h_{\mu\nu}(\epsilon\zeta,{\bm x})}\Bigr]_{{\bm x}={\bm x}_N(\zeta)}\;.
\end{equation}
We emphasize that in order to integrate the right side of Eq. (\ref{aa}) one, first, has to take the derivative and, then, to substitute the unperturbed light ray trajectory for ${\bm x}$. 

One can prove that 
\begin{equation}
\label{dd}
c\,k^\alpha\Bigl[\partial_\alpha h_{\mu\nu}(\epsilon t,{\bm x})\Bigr]_{{\bm x}={\bm x}_N(t)}={d\over dt}\biggl[h_{\mu\nu}(\epsilon t,{\bm x}_N(t))\biggr]\;,
\end{equation}
where $d/dt$ denotes the total time derivative along the unperturbed light ray. By making use of Eq. (\ref{dd}) one can perform one integration in Eq. (\ref{aa}) and recasts it to a simpler form
\begin{equation}
\label{ty}
\Delta(t_1,t_0)={1\over 2}\,k^\mu k^\nu \int_{t_0}^{t_1}h_{\mu\nu}\left(\epsilon t,{\bm x}_N(t)\right)\,dt \;.
\end{equation}
This equation must be solved for $\epsilon\not=1$. It will show at what terms the speed of gravity appears explicitly. Then, we take a general relativistic limit $\epsilon\rightarrow 1$. This is a way to distinguish the effects associated with the speed of gravity (dynamics of gravitational field) from those related to the process of light propagation.

Relativistic time delay is found by performing integration in Eq. (\ref{ty}) with positions of the gravitating bodies in the metric tensor (\ref{mt1})--(\ref{mt3}) treated as functions of time ${\bm x}_a(\tau)={\bm x}_a(\epsilon t)$. The integral (\ref{ty}), however, is too complicated if the true orbital motion of gravitating bodies is used and its exact analytic calculation can not be done explicitly. For this reason, we approximate the integral by making use of a linear Taylor expansion
\begin{eqnarray}
\label{v1}
x^i_a(\tau)&=&x^i_a(\tau_a)+v^i_a(\tau_a)(\tau-\tau_a)+O\left(\tau-\tau_a\right)^2\;,\\\nonumber\\
\label{v2}
v^i_a(\tau)&=&v^i_a(\tau_a)+O\left(\tau-\tau_a\right)\;,
\end{eqnarray}
where $\tau=\epsilon t$ and $\tau_a$ is an arbitrary instant of time that can not be unambigiously determined in the post-Newtonian calculations. In order to fix $\tau_a$ one needs to resort to the post-Minkowskian result of integration of the light-ray propagation equation obtained in \cite{2}. This is a very subtle point closely related to the interpretation of the relativistic VLBI experiment. It is intuitievly clear that $\tau_a$ should be close to the time of observation but its exact value can be found only in the framework of the exact post-Minkowskian theory of integration of equations of light, propagating in variable gravitational fields \cite{2}. We shall show how to fix the value of $\tau_a$ in the next section. 

We also emphasize that ${\bm x}_a(\tau_a)$ and ${\bm v}_a(\tau_a)$ are coordinate and velocity of the $a$th gravitating body taken at the time $\tau_a$ on its real trajectory. Hence, if time $\tau$ in the expansions (\ref{v1}), (\ref{v2}) is significantly different from $\tau_a$, the position ${\bm x}_a(\tau)$ of the body approximated on the basis of these expansions can be very far away from its real position at the time $\tau$. This indicates to the limitation of the post-Newtonian integration technique which can be cured only within the post-Minkowskian approach \cite{5}, \cite{km}.  

Substituting Eqs. (\ref{und}), (\ref{v1}) and (\ref{v2}) for ${\bm x}$, ${\bm x}_a(\tau)$ and ${\bm v}_a(\tau)$ in the metric perturbations (\ref{mt1})--(\ref{mt3}) and integrating, we obtain the relativistic time delay 
\begin{equation}
\label{5aa}
\Delta(t_1,t_0)=-2\sum^N_{a=1}\frac{GM_a}{
c^3p_a}\left(1-\frac{2}{c_g}{\it\bm k}{\bm\cdot}{\it\bm v}_a\right)
\ln\left(p_a R_{1a}-{\it\bm p}_a{\bm\cdot}{\it\bm R}_{1a}\right)
+C_1\left({\it\bm k},{\bm x}_0\right)
\,,
\end{equation}
where $p_a=|{\it\bm p}_a|$, ${\it\bm p}_a\equiv{\it\bm k}-{\it\bm v}_a/c_g$ \footnote{The speed of gravity $c_g$ appears in the expression for the vector ${\bm p}_a$ because one has ${\bm x}(t)-{\bm x}_a(\tau)\sim c{\bm k}t-\epsilon{\bm v}_at=c{\bm p}_a t$. From the physical point of view it means that motion of light and the bodies in the gravitating system goes on different time scales if $c_g\not= c$.}, 
 ${\it\bm R}_{1a}\equiv{\it\bm x}_1-{\it\bm x}_{a}(\tau_1)$, $R_{1a}=|{\it\bm R}_{1a}|$, $\tau_1=\epsilon t_1$, ${\it\bm x}_1\equiv{\it\bm x}(t_1)$, ${\bm x}_a(\tau_1)$ and ${\bm v}_a$ must be understood as given by Eqs. (\ref{v1}), (\ref{v2}).  
 
 The constant term $C_1\left({\it\bm k},{\bm x}_0\right)$ is the value of the integral in Eq. (\ref{ty}) taken at time $t_0$ when light was emitted by the quasar. It has no real physical meaning in the post-Newtonian approach and could violate consistency of the calculation for close celestial objects being members of our galaxy. However,  the quasar in the relativistic VLBI experiment on September 8, 2002 is lying extremaly far away, so that the constant $C_1$ is unimportant in the following discussion because it can be ruled out in calculation of the differential VLBI time delay that is a real observable quantity. Physically meaningful value of the constant of integration $C_1$ can be also obtained only by making use of the post-Minkowskian approach \cite{2}, \cite{5}, \cite{km}.

\section{The Retardation of Gravity}

By making use of the post-Minkowskian theory of light propagation we have predicted \cite{2} that the VLBI experiment on September 8, 2002 is able to measure the speed of gravity by observing the effect of the {\it retardation of gravity} propagating from the sun-orbiting Jupiter to light ray. In this section we confirm this interpretation on the basis of the post-Newtonian approach. To this end, let us recast Eq. (\ref{5aa}) to another form by making use of the Taylor expansion of vector ${\bm p}_a$ with respect to the small parameter $v_a/c_g$. This yields
\begin{equation}
\label{6aa}
\Delta(t_1,t_0)=-2\sum^N_{a=1}{GM_a\over
c^3}\left(1-{1\over c_g}{\it\bm k}{\bm\cdot}{\it\bm v}_a\right)
\ln\Bigl[p_a\left(R_{1a}-{\it\bm n}_a{\bm\cdot}{\it\bm R}_{1a}\right)\Bigr]
+C\left({\it\bm k},{\bm x}_0\right)\,,
\end{equation}
where $C\left({\it\bm k},{\bm x}_0\right)$ is (another) constant of integration and the unit vector \footnote{We again emphasize that the velocity ${\bm v}_a\equiv{\bm v}_a(\tau_a)$ is the physical velocity of the $a$th body. } 
\begin{equation}
\label{aber}
{\it\bm n}_a={\it\bm k}-{1\over c_g}{\it\bm k}\times({\it\bm v}_a\times{\it\bm k})
\end{equation}
explicitly describes a change in the direction of propagation of light that can be interpreted as {\it dragging of the light-ray} caused by the flux of the gravitational field generated by the translational motion of $a$th body. This phenomena can be equivalently interpreted as the effect of aberration of gravity, i.e. displacement of the observed body from its present to retarded position, which is unobservable in the equations of motion of slowly-moving particles \cite{carlip} but can be detected in the light-ray deflection relativistic experiments in time-dependent gravitational fields. 

This can be demonstarted if one chooses the time $\tau_a$ in Eqs. (\ref{v1}) and (\ref{v2}) to be coinciding with the retarded time $s_1$ corresponding to the instant of observation $t_1$ and reflecting the fact that gravity takes finite time to propagate from the moving gravitating body to observer. This time is calculated from Eq. (\ref{rt}) as follows
\begin{equation}
\label{rt1}
s_1=\tau_1-{r_{1a}(s_1)\over c_g}\;,
\end{equation}
where $\tau_1=\epsilon t_1$, ${\bm r}_{1a}(s_1)={\bm x}_1-{\bm x}_a(s_1)$, and $r_{1a}(s_1)=|{\bm r}_{1a}(s_1)|$. The geometric distance ${\bm R}_{1a}$ between the observer and the $a$th body reads
\begin{eqnarray}\label{pp}
{\bm R}_{1a}&=&{\bm r}_{1a}(s_1)-{r_{1a}(s_1)\over c_g}\,{\bm v}_a(s_1)+O(c_g^{-2})\;,\\
\label{qq}
R_{1a}&=&r_{1a}(s_1)-{1\over c_g}\,{\bm v}_a(s_1){\bm\cdot}{\bm r}_{1a}(s_1)+O(c_g^{-2})\;.
\end{eqnarray} 
Substituting Eqs. (\ref{pp}) and (\ref{qq}) to Eq. (\ref{6aa}) and reducing similar terms one obtains
\begin{equation}
\label{6bb}
\Delta(t_1,t_0)=-2\sum^N_{a=1}\frac{GM_a}{
c^3}\Bigl(1-\frac{1}{ c_g}{\it\bm k}{\bm\cdot}{\it\bm v}_a\Bigr)
\ln\Bigl[r_{1a}(s_1)-{\it\bm k}{\bm\cdot}{\it\bm r}_{1a}(s_1)\Bigr]
+C\left({\bm k},{\bm x}_0\right)\,.
\end{equation}
This formula gives the post-Newtonian time delay in terms of the retarded position of the solar system bodies and coincides in the limit $c_g\rightarrow c$ with the analogous formula derived by \cite{2} on the basis of the post-Minkowskian approach \footnote{Let us emphasize that if one did not choose $\tau_1=s_1$ such a coincidence could not be achieved.}. It is evident from Eq. (\ref{rt1}) that the retardation in Eq. (\ref{6bb}) is due to the finite speed of propagation of gravity $c_g$
because Eq. (\ref{rt1}) is just the equation of the gravity null cone describing propagation of gravity between the moving gravitating body and the point of observation. Relativistic VLBI experiment on September 8, 2002 measured the retarded position of Jupiter in the sky through the gravitational deflection of light of the quasar which was in the plane of the sky at the distance of $\sim 14$ jovian radii from Jupiter (see Fig. \ref{fig2}).

\section{Discussion}

We have constructed a speed-of-gravity parameterization of the linearized equations of general relativity and calculated a speed-of-gravity correction to the Shapiro time delay given by Eq. (\ref{6bb}). Its post-Newtonian expansion is identical with Eq. (12) from our previous work \cite{2} where we used the post-Minkowskian approach and introduced the fitting parameter $\delta$ on a phenomenological basis. The present paper proves that on the sequence of the parametric space-times manifolds, $\delta=c/c_g-1$ and distinguishes two fundamental speeds in the Einstein and Maxwell equations. Hence, the relativistic VLBI experiment conducted on September 8, 2002 allows to determine the speed of gravity and to measure its numerical value.

Einstein theory of general relativity predicts that perturbation of the gravitational field produced by orbital motion of Jupiter propagates with the speed of gravity $c_g=c$ towards the position of the light-ray particle traveling from quasar to Earth. This prediction can be tested by VLBI technique without direct detection of gravitational waves. The VLBI experiment gives the magnitude of the retardation-of-gravity effect in terms of its speed $c_g/c=1.06\pm 0.21$ \cite{apj}. We conclude that the speed of gravity is the same as the speed of light within our experimental error.

\subsection{Acknowledgments}

We are grateful to B. Mashhoon, G. Sch\"afer, B. Iyer, I.I. Shapiro, A. Aliev and M. Lattanzi for fruitful and stimulating discussions as well as for criticism of this work. My special thanks go to Dr. Henry W. White for his support of this research as a chair of the Department of Physics and Astronomy of the University of Missouri-Columbia. A part of this work was supported by the Feza G\"ursey Institute (Istanbul, Turkey) and Torino Astronomical Observatory (Italy) travel grants. The Epply Foundation for Research Award 002672 was essential in pursuing this advanced research. 

\appendix
\section{Comments on the paper by Asada \cite{1}}
Asada \cite{1} calculated the relativistic time delay which formally coincides with our formula (\ref{6bb}). Asada's calculations are not complete as he {\it postulated} in his calculations that position of Jupiter must be fixed at the retarded instant of time $s_1=t_1-(1/c)|{\bm x}_1-{\bm x}_J(s_1)|$. Such postulate is a plausible {\it assumption} but not a mathematical proof. For this reason Asada's interpretation of the relativistic VLBI experiment on September 8, 2002 is ungrounded.

 In our previous paper \cite{2} we have solved the system of the Einstein and light geodesic equations without making any restrictive assumptions about the motion of Jupiter and {\it proved} that the position of Jupiter must be taken at the retarded instant of time $s_1$ which is an argument of the metric tensor and, hence, describes the effect of propagation of gravity with the gravity speed $c_g=c$.  

\section{Comments on the paper by Will \cite{w-astro}}\label{a2}

Will \cite{w-astro} calculated the relativistic time delay by making use of the matched asymptotic technique of the solutions of the light geodesic equation solved in the near- and far-zone fields of the gravitational potentials $U$ and $V_i$ obtained from the field equations 
\begin{eqnarray}
\label{wi1}
\left(-{1\over c_g^2}{\partial^2\over\partial
t^2}+{\bm\nabla}^2\right)U(t, {\bm x})&=&-{4\pi G\over
c^2}\rho(t, {\bm x})\;,\\\nonumber\\\label{wi2}
\left(-{1\over c_g^2}{\partial^2\over\partial
t^2}+{\bm\nabla}^2\right)V_i(t, {\bm x})&=&-{4\pi G\over
c^3}\rho(t, {\bm x})v^i(t, {\bm x})\;,
\end{eqnarray}
where $\rho(t, {\bm x})$ is the density of matter and the time $t$ is the same as in the light geodesic equations. 

Eqs. (\ref{wi1}), (\ref{wi2}) are not Einstein equations and are written in a particular gauge which was not clearly described. Furthermore, we notice that the field equations (\ref{wi1}), (\ref{wi2}) are, in fact, disconnected from the light geodesic equations. Indeed, assuming that $c_g\not=c$ and using a single time scale $t$ imply that vacuum must be considered as having a "refractive" index $n\sim c/c_g\not=1$ and the equations of light propagation should have extra terms proportional to $n-1$ \cite{synge} that were not discussed in \cite{w-astro}. The whole problem of the unambiguous interpretation of the VLBI experiment under discussion is, in fact, a matter of construction of a self-consistent bi-metric theory of gravity and electromagnetism. The PPN formalism \cite{9} can not play a role of such a theory as it consists of a set of plausible but, in any other aspect, disconnected assumptions about how gravity and electromagnetic fields behave if general theory of relativity is violated. 

\begin{figure}
\centerline{\psfig{figure=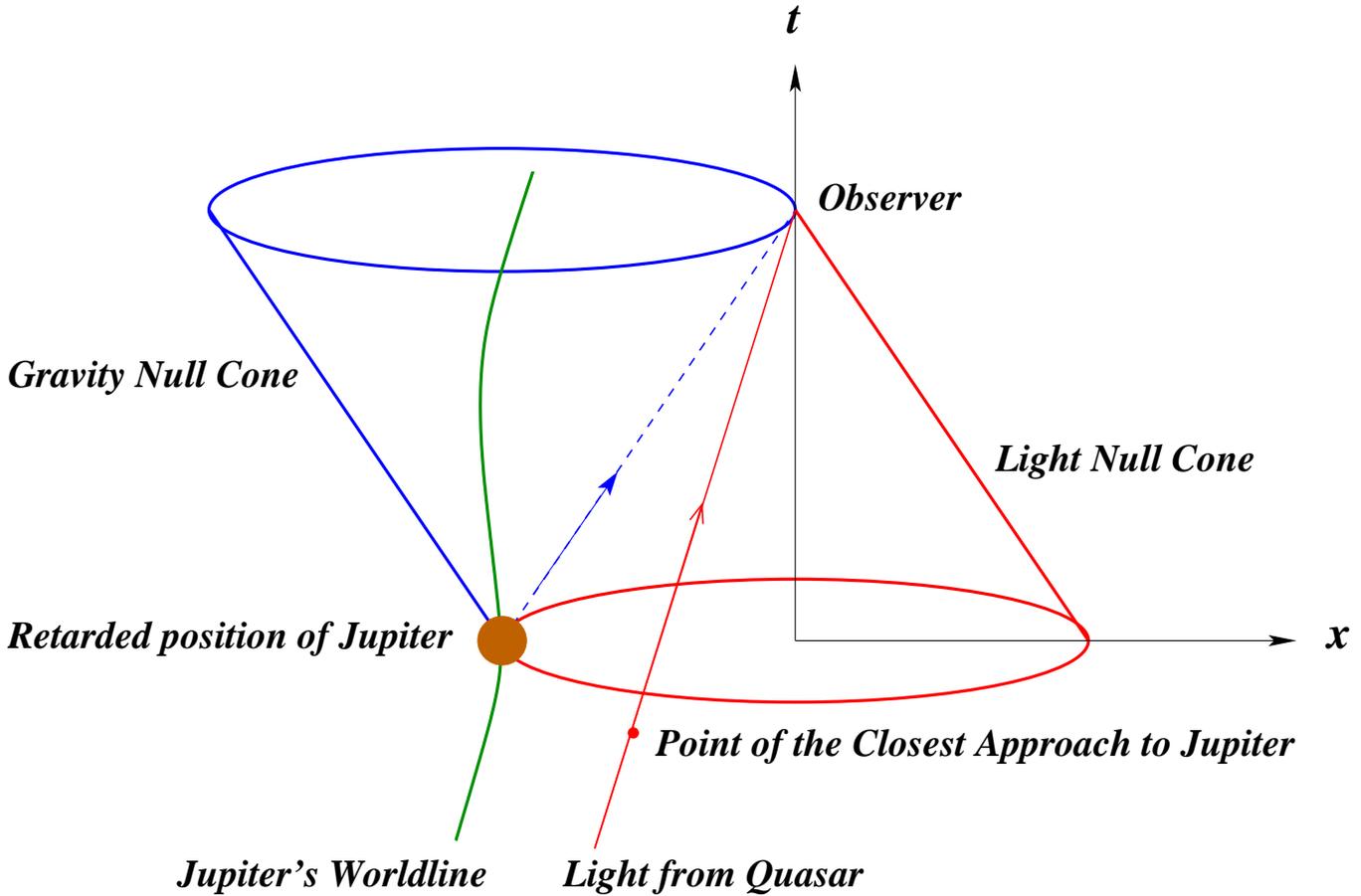,angle=0,height=12cm,width=18cm}}\vspace{1cm}
\caption{Minkowski diagram showing the relative configuration of gravity and light null cones related to the point of observation. Two cones do not intersect but are tangent to each other 
along a single (dashed) line. The tangent line of the two null cones connects Jupiter and observer. Light from quasar propagates 
along a different world line on the light null cone emanating from the quasar (not shown). The gravitational interaction between moving Jupiter and a light particle always occurs along the gravity null cone. 
In other words, when Jupiter moves it continuously sends gravitational ``messages" about its 
position in space that propagate along the gravity null cone  with the finite speed $c_g=c$ towards the light particle. The last such ``message" reaches the light particle at the point of observation and points out to  
the retarded position of Jupiter. This interplay between gravity and light was used in the VLBI experiment on September 8, 2002 to test how fast gravity propagates and to measure its speed \cite{2}, \cite{apj}. Will's statement \cite{w-astro} is that position of Jupiter in the VLBI time delay formula (\ref{6bb}) must be taken at the time of the closest approach of light to Jupiter. \label{fig2}}
\end{figure}

There are two other concerns about validity of the calculations given by Will \cite{w-astro}. The first concern relates to the meaning of the retarded time $s_a=\sigma-r_a(\sigma,s_a)/c_g$ in the arguments of the bodies ${\bm x}_a(s_a)$. Will insists (see appendix of \cite{w-astro}) that coefficients of the Taylor expansion ${\bm x}_a(s_a)={\bm x}_a(0)+{\bm v}_a(0)s_a+...$ must be understood as taken at the time $\sigma=0$. This contradicts to the definition of the Taylor expansion given in any mathematical textbook. By definition of the Taylor expansion the coefficients ${\bm x}_a(0)$ and ${\bm v}_a(0)$ in the formula given above are taken at the retarded time $s_a=0$ \footnote{Misinterpretation of the retarded time arguments led Will to the erroneous use of the Taylor expansion in his Eq. (34) that, as a matter of fact, must be used along with the gravity null cone equation (18).}. This instant of time can coincide with that $\sigma=0$, if and only if, the retarded time equation $s_a=-r_a(0,s_a)/c_g$ has a solution $s_a=0$. From the astronomical point of view it means that the light ray must pass through the light-ray deflecting body. This is indeed coincides with what is shown in the figure of Will's paper. However, in the relativistic VLBI experiment on September 8, 2002 light was passing pretty far away from Jupiter at the distance of about $14$ jovian radii (see Fig. \ref{fig2}). Hence, for $\sigma=0$ the time $s_a\not=0$ in the VLBI experiment.

The second concern relates to the definition of the point of the closest approach of the light ray to the moving body. After removing the misinterpretation of the Taylor expansion in \cite{w-astro} one has to understand that Will has {\it postulated} that the point of the closest approach is reached by light at the instant of the retarded time $s_a=0$. Will also {\it postulated} that at the point of the closest approach the light-ray-propagation unit vector ${\bm k}$ and that of the impact parameter ${\bm d}_a(s_a)={\bm\xi}-{\bm x}_a(s_a)$ must be orthogonal in the Euclidean sense. We emphasize that the point of the closest approach of light ray to the body must be found after solution of the corresponding equation giving the distance between the body and the light particle as a function of time, let say, $D(\sigma,s_a)=|c{\bm k}\sigma+{\bm d}_a(s_a)|$. Finding the minimum of this function from the condition $dD(\sigma,s_a)/d\sigma=0$, shows that at the point of the closest approach the two vectors ${\bm k}$ and ${\bm d}_a(s_a)$ are not orthogonal and their Euclidean dot product is proportional to the relativistic terms of order $d_a v_a/c_g$ which are important in the subsequent calculations of the time delay. All such terms are absent in \cite{w-astro} that along with the previous our comments casts strong doubts on the validity of Will's calculations and his interpretation of the relativistic VLBI experiment conducted by Fomalont and Kopeikin \cite{apj}.

\newpage
\newpage

\end{document}